\documentclass[singlespacing]{elsart}

\usepackage{graphicx}

\usepackage{amssymb}
\journal{}
\begin{document}

\begin{frontmatter}

\title{Model for solvent viscosity effect on enzymatic reactions}

\author{A.E. Sitnitsky},
\ead{sitnitsky@mail.knc.ru}

\address{Institute of Biochemistry and Biophysics, P.O.B. 30, Kazan
420111, Russia. e-mail: sitnitsky@mail.knc.ru }

\begin{abstract}
Why reaction rate constants for enzymatic reactions are typically inversely proportional to fractional power exponents of solvent viscosity remains to be already a thirty years old puzzle.
Available interpretations of the phenomenon have not led to consensus among researches about its origin. They invoke to either a modification of 1. the conventional Kramers' theory  or that of 2. the Stokes law. We show that there is an alternative interpretation of the phenomenon at which neither of these modifications is in fact indispensable. Basing on an analogy from the theory of adsorption on heterogeneous surfaces we reconcile 1. and 2. with the experimentally observable dependence. We assume that an enzyme solution in solvent with or without cosolvent molecules is an ensemble of samples with different values of the viscosity for the movement of the system along the reaction coordinate.
We assume that this viscosity consists of the contribution with the weight $q$ from cosolvent molecules and that with the weight $1-q$ from protein matrix and solvent molecules. We introduce heterogeneity in our system with the help of a distribution over the weight $q$. The function of the distribution is a unique characteristic of the solution of enzymes (type of the enzyme, cosolvent molecular weight, pH, temperature, etc.). We verify the obtained solution of the integral equation for the unknown function of the distribution by direct substitution.
We conclude that even at linear relationship between the solvent viscosity and that for the movement of the system along the reaction coordinate our approach enables us to obtain the required dependence. All parameters of the model are related to experimentally observable values.
The meaning of fractional exponents appears to be the characteristic for the behavior of the distribution with the variation of the weight $q$. Our approach yields the existence of the limit value for the fractional power exponent with the decrease of cosolvent molecular weight that is in agreement experimental data known from the literature. This limit value is determined by the properties of the protein structure and thus is a unique characteristic of the type of the enzyme only rather than that of the solution. General formalism is exemplified by the analysis of literature experimental data for oxygen escape from hemerythin.
\end{abstract}

\end{frontmatter}

\section{Introduction}
Viscosity dependence of enzymatic and protein (ligand binding/rebinding) reactions
is a long standing problem for biophysics \cite{Gav78},
  \cite{Gav79}, \cite{Bee80}, \cite{Gav80}, \cite{Dos83}, \cite{Gav86},
 \cite{Fra88},  \cite{Dem89}, \cite{Ng91},
\cite{Ng911}, \cite{Gav94}, \cite{Dos94}, \cite{Yed95}, \cite{Bar95}, \cite{Oh97}, \cite{Kle98},
\cite{Fra99}, \cite{Uri03}, \cite{Sit08}. For such reactions the functional dependence of the
reaction rate constant for the rate
limiting stage $k$ on solvent viscosity $\eta$ has the form
\begin{equation}
\label{eq1} k\propto \frac{1}{\left (\eta/\eta_0 \right )^{\beta}}
\end{equation}
where $\eta_0$ is the viscosity of pure solvent (for water $\eta_0 = 1\ cP$ at room temperature) and
$ 0 < \beta < 1 $ (usually $\beta \approx 0.4 \div 0.8$). This dependence is experimentally verified
in the range of variation of solvent viscosity by two orders of magnitude $\eta < 100\ cP$.
Similar dependence also takes place for folding of proteins
(see \cite{Pab04}, \cite{Fra06}, \cite{Kum08} and refs. therein) and at formation of protein structure
\cite{Jas01}. However we will not touch upon these processes in the present paper.

Much efforts were devoted to explaining the functional dependence (\ref{eq1}) in the previous
century with no obvious consensus and commonly accepted mechanism of the phenomenon. For instance
the authors of \cite{Yed95} conclude that "there seems to be no general
agreement yet about the origin of the fractional $\beta$ value in Eq.1".
The authors of \cite{Kle98} draw to a similar conclusion: "At present there is no general
agreement on the meaning of fractional exponents implying
a breakdown of Stokes law". Despite steady growth of experimental data
little has changed in this issue since the date of the cited papers (cf., e.g.,:
"The meaning of fractional exponents
implying that the friction coefficient does not vary linearly with
the solvent viscosity, essentially a violation of Stokes law ($f \propto
\eta_s$), is not clear at present" \cite{Kum08}).
Detailed studies revealed that in fact the fractional
index of a power $\beta$
is a function of cosolvent molecular weight $M$ (i.e., the mass of
a cosolvent molecule expessed in atomic units and measured in Daltons) $\beta=\beta(M)$ \cite{Yed95}.
If one varies the solvent viscosity by large cosolvent molecules with high
molecular weight that do not penetrate into enzyme then one obtains that the
fractional exponent $\beta \rightarrow 0$, i.e., the reaction rate constant does
not depend on solvent viscosity. With the decrease of cosolvent molecular
weight the fractional exponent $\beta$ increases. In the limit of
hypothetical "ideal" cosolvent with very small molecular weight
(cosolvent molecules freely penetrate into enzyme and are distributed
there homogeneously) it tends to some limit value
\begin{equation}
\label{eq2} \beta_{max}=\lim_{M\to M_{min}}{\beta(M)}
\end{equation}
For oxygen escape from hemerythin this value is $\beta_{max}\approx 0.79$.
The latter is neither experimental value nor a calculated one.
It is an extrapolated number (see \cite{Yed95} for details). The existence of enzymatic reactions with
$\beta =1$ \cite{Fra99}, \cite{Fra06} suggests that the limit value is a unique characteristic of the type of the enzyme and can take the values up to $1$ as well.

The main tool to describe the viscosity dependence of a reaction rate constant is the high friction limit
(also called strong damping or overdamped regime) of the Kramers' theory  \cite{Han90}. In it the reaction is conceived as a diffusion process of a particle with some effective mass along a reaction coordinate over some potential surface. The friction coefficient for the particle $\nu$ is supposed to obey the Stokes law $\nu \propto \eta$ that yields the well known dependence
\begin{equation}
\label{eq3} k_K\propto \frac{1}{\eta/\eta_0}
\end{equation}
As was stated above for enzymatic reaction this dependence is inconsistent with observations.
There are two strategies to interpret the experimentally observed dependence (\ref{eq1}) at present:
1. to modify the Kramers' theory and 2. to modify the Stokes law.
The first approach usually leads to rather complicated theoretical constructions that have not proved to yield a universal and commonly accepted resolution of the problem.
It is realized in, e.g., the Grote-Hynes theory (GH) \cite{Gro80}, the Zwanzig model  \cite{Zwa92} or the model suggested in \cite{Sit06}, \cite{Sit08}. GH gives that the rate dependence on solvent viscosity should be weaker than that predicted by Kramers' one.
The authors of this model argue that the friction coefficient should
be proportional to a high-frequency viscosity at an appropriately
renormalized frequency \cite{Gro80}. Assuming a positive power dependence
of viscosity on frequency, they find that the renormalized high frequency
viscosity has a fractional power dependence on the
low-frequency viscosity which can be measured as the usual
viscosity $\eta$. Thus their assumption is in fact introducing the fractional exponent "by hand" in a phenomenological manner.
The Zwanzig model infers the reaction rate constant from the first principles but yields too small value for the fractional exponent $\beta=0.5$. The model \cite{Sit06}, \cite{Sit08} deals only with the
limiting case of the hypothetical "ideal" cosolvent with very small molecular weight
and yields for the fractional exponent
the limit value $\beta_{max}\approx 0.75$ that is in good agreement with the
extrapolated one  $\beta_{max}\approx 0.79$  from the paper \cite{Yed95}. No extension of this model
to the case of realistic cosolvent with finite molecular weight has been suggested yet.
The second approach takes into account that the value of viscosity in enzyme active site differs from that of the solvent \cite{Gav79} and may be a {\it nonlinear} function of the latter.
It is  suggested that the fractional exponent $\beta$ is the degree with which solvent viscosity is coupled with (frequency dependent friction) \cite{Dos83}, \cite{Sch88}, \cite{Kle98},
or penetrates into  (position dependent friction) \cite{Gav80},
\cite{Bar95} the protein interior. Any of these variants yields the modification of the Stokes law
$\nu \propto \eta^\beta$. However the fractional exponent $\beta$ in these approaches still appears as an empirical parameter. It accounts for the nonlinearity of the relationship between the values of the solvent viscosity and that in the enzyme active site but its value is not calculated and its origin remains mysterious.

In our opinion there is another way to interpret the experimental data that requires neither modification of the Kramers' model nor that of the Stokes law. Thus
the aim of the present paper is to reconcile the conventional Kramers' model and the Stokes law with the experimentally observed dependence (\ref{eq1}). The main premise of our approach is in the fact that a realistic enzyme solution in solvent with or without cosolvent molecules is actually an ensemble of enzymes with different conditions (values of viscosity for the movement of the system along the reaction coordinate).
We show that even at {\it linear} relationship between solvent viscosity and that for the movement of the system along the reaction coordinate we can obtain (\ref{eq1}) if we take into account heterogeneity of conditions in the ensemble. It should be stressed that the present approach can be generalized to be based not on the Kramers' model but on the models \cite{Gro80}, \cite{Zwa92} or  \cite{Sit06}, \cite{Sit08}. However for the sake of definiteness we choose the Kramers' model and reconcile the simple dependence (\ref{eq3}) with the experimentally observable expression for the reaction rate constant (\ref{eq1}).
The latter is obtained by averaging of individual Kramers' rate constants over the distribution.
The aim of the present paper is to show that the idea of heterogeneity
enables one to resolve the thirty years old puzzle of solvent viscosity effect on enzymatic reaction rate constant
in a conceptually much more simple way than modification of either the Kramers' theory or that of the Stokes law. Also we stress that our distribution has nothing to do with the
nonthermally equilibrium form of the reactant distribution during the reaction employed in the approach of \cite{Sum91}.

There is a noteworthy analogy between the problem under consideration and that of adsorption on heterogeneous surfaces. In the latter one reconciles the  Langmuir isotherm grounded by statistical mechanics and giving the number of adsorbed molecules $\theta$ proportional to pressure $P$ in the low pressure limit $\theta \propto P$ with the
experimentally observable phenomenological Freundlich isotherm giving the dependence $\theta \propto P^\delta$  where $0 < \delta <1$. One attains this aim
via introducing a distribution over energy due to heterogeneity of
the surface. As is well known the surface heterogeneity in adsorption theory is
taken into account by integral equation approach (see, e.g., \cite{Rud92},
\cite{Cer93} and refs. therein). This line in adsorption
theory has a long history. It was initiated by the paper of  Zel'dovich
\cite{Zel34} (see, e.g., \cite{Tho67} for authoritative discussion of priority
questions). The next important steps were made by Sips \cite{Sip48}, \cite{Sip50}
and Misra \cite{Mis70} who used Stieltjes transform. Detailed discussion of
related problems is given in \cite{Cer72}, \cite{Rud75}, \cite{Jar75}.
For the solution of the corresponding integral equations the authors of these papers used the so
called condensation approximation \cite{Cer72}, \cite{Cer93}.
Finally this line was culminated by the paper of Landman and Montroll \cite{Lan76}
who applied powerful Wiener-Hopf technique for the solution of the corresponding
integral equations. The enormous preceding experience accumulated in this field of chemical physics provides an invaluable source of information for our problem both from the side of conceptual aspects and that of mathematical technique.

The paper is organized as follows. In Sec. 2 the character of the relationship between
the viscosity for the movement of the system along the reaction coordinate
and that of solvent viscosity is discussed. In Sec. 3 we formulate the integral equation
for taking into account the heterogeneity of conditions in the enzymes from the ensemble via introducing the distribution over the weight $q$. In Sec. 4 we cast this equation to the form solvable with the help of Fourier transform and verify our solution by direct substitution. In Sec.5 we obtain the relationship of the all parameters of the model with experimentally observable values.
In Sec. 6 the results are discussed and the conclusions are summarized.
In Appendix some known mathematical formulas are collected for convenience.

\section{Viscosity for the movement of the system along the reaction coordinate}
The reaction coordinate in the Kramers' theory is a notion that enables one to reduce very complicated dynamics of the system in a multidimensional configurational space to the movement of a particle with some effective mass along one effective dimension. For the case of an enzymatic reaction the movement of the system along the reaction coordinate is determined by: processes in the substrate molecule (e.g., stretching of the bond to be cleaved); friction for these processes by the environment in the active site of the enzyme (that is stipulated by solvent and cosolvent molecules and also protein side chains located there); relevant motion for catalysis of side chains and larger fragments of structure from protein interior (such as, e.g., $\alpha$-helicies and $\beta$-sheets) and friction to it by solvent and cosolvent molecules and also by protein matrix in their environment; etc. We denote the viscosity for the movement of the system along the reaction coordinate as $\zeta$ and suppose the conventional Stokes law for the friction coefficient of the particle $\nu \propto \zeta$. The solvent viscosity we denote as earlier $\eta$. The linear relationship between $\eta$ and $\zeta$ can be motivated as follows.

There seems to be consensus among researches that the effect of solvent viscosity on the enzymatic reaction is mediated by protein dynamics
\cite{Gav79}, \cite{Bee80}, \cite{Dos83}, \cite{Gav86},
\cite{Dem89}, \cite{Ng91},
\cite{Ng911}, \cite{Gav94}, \cite{Dos94}, \cite{Oh97}.
That is why we further identify the viscosity for the movement of the system along the reaction coordinate with the viscosity at relevant motion for catalysis of individual fragments in protein interior and in enzyme active site.
Experiment yields that protein dynamics over a wide range of temperatures obey the relationship between solvent viscosity $\eta$ and the mean square displacement (MSD) $<u^2>$ \cite{Mag04}, \cite{Cor05}
\begin{equation}
\label{eq4} <u^2>=\frac{g}{\ln \left(\eta/\eta_0\right)}
\end{equation}
where $g$ is a constant.
We notice that firstly (\ref{eq4}) is divergent at $\eta/\eta_0 =1$ and secondly nobody verifies (\ref{eq1}) experimentally
in the limit $\eta/\eta_0 \rightarrow 1+$. In practice one starts from some values noticeable larger than $1$. For instance in \cite{Yed95} $\left (\eta/\eta_0\right)_{min}\approx 1.6$. That is why further on we consider the range of solvent viscosity
\begin{equation}
\label{eq5} \frac{\eta}{\eta_0} \geq 2
\end{equation}
The theoretical counterpart of (\ref{eq4}) given by, e.g., the most simple and extraordinary successful "model of bounded diffusion"
\cite{McC77}, \cite{Kna82}, \cite{Dos83} yields for the MSD the expression
\begin{equation}
\label{eq6} <u^2>=\frac{k_B T}{m\omega_0^2}
\end{equation}
where $k_B$ is the Boltzman constant, $T$ is the temperature, $m$ is the mass of the moving fragment and $\omega_0$ is the frequency defined by the curvature of the potential well near its bottom for the motion of the fragment. This motion is determined by the value of the local viscosity $\zeta$ of the environment for the motion of the fragment in protein interior.

First we notice that a nonlinear relationship between $\zeta$ and $\eta$ is compatible with (\ref{eq4}) and (\ref{eq6}) with the linear one being a particular case.
We assume that $\zeta/\eta_0$ consists of the contribution with the weight $1-q$ determined by protein matrix and solvent molecules
and the relative contribution with the weight $q$ from cosolvent molecules in the fragment environment.
The weight $q$ may take values from the range
\begin{equation}
\label{eq7} 0 < q < 1
\end{equation}
The term "relative" means that the latter contribution must tend to zero for the case of pure solvent (without cosolvent molecules). That is why
for it we adopt the conventional Andrade-Eyring-Frenkel' form $\exp\left[E/\left (k_BT\right )\right]$ subtracted by $1$ (because for the case of pure solvent $\eta/\eta_0=1$ or equivalently at $E=0$ the exponent is $1$). Here $E$ is the activation energy ($E \geq 0$). Thus we consider the form
\begin{equation}
\label{eq8} \zeta/\eta_0=a\left \lbrace(1-q)+q\left[\exp\left(\frac{E}{k_BT}\right)-1\right]\right \rbrace
\end{equation}
The parameter $a$ is related to the fractional power exponent $\beta$ that will be obtained later on theoretical grounds. Also below the  explicit dependence $a=a\left(M\right)$ of the parameter on the cosolvent molecular weight $M$ will be obtained from the analysis of experimental data.

One can easily see that the nonlinear relationship between the solvent viscosity $\eta$ and the local viscosity for the motion of the fragment in protein interior $\zeta$ of the form
\begin{equation}
\label{eq9} \frac{\zeta}{\eta_0}=a\left \lbrace(1-q)+q\left[\left(\frac{\eta}{\eta_0}\right)^\kappa-1\right]\right \rbrace
\end{equation}
is compatible with the relationships (\ref{eq4}), (\ref{eq6}) and (\ref{eq7})
at any $\kappa > 0$ if we set
\begin{equation}
\label{eq10} g=\frac{E}{m\omega_0^2\kappa}
\end{equation}
At $\kappa=1$ we have the particular linear case
\begin{equation}
\label{eq11} \frac{\zeta}{\eta_0}=a\left [(1-q)+q\left(\frac{\eta}{\eta_0}-1\right)\right ]
\end{equation}
There in no wonder that we can obtain the experimentally observable dependence (\ref{eq1}) for the nonlinear case $0 < \kappa < 1$ simply by setting $\beta = \kappa$. Our aim is to stress that even the linear case $\kappa =1$ can lead to the required dependence (\ref{eq1}) as well. At the same time it should be emphasized that the assumption about linearity is by no means crucial for the validity of our approach. The latter can still remain workable for the general nonlinear case even at $\kappa \not= \beta$. However we will not explore this option in the present paper and restrict ourselves by the linear case $\kappa =1$. In our opinion namely this case can elucidate the essence of our approach with full clarity. We pursue this line in the next Sec.

\section{Formulation of the integral equation}
The value of viscosity for the movement of the system along the reaction coordinate
may be different for different enzyme molecules (samples) from the ensemble of enzymes.
In the individual samples from the ensemble we have heterogeneity of conditions both in the enzyme active sites and in protein interior.
That is why the experimentally observable dependence of the reaction rate constant on solvent viscosity should be determined by some averaged value of the viscosity for the movement of the system along the reaction coordinate. We consider the following way to take into account this heterogeneity by averaging with the help of some integral equation.

We assume that due to heterogeneity of conditions in the individual samples from the ensemble of enzymes
the weight $q$ in (\ref{eq8}) and consequently in (\ref{eq11}) can take different values from the range $0 < q < 1$.
Thus the ensemble is characterized by the distribution $\rho (q)$ over the values of the weight $q$.
We also assume that the reaction rate constant for the individual sample from the ensemble of enzymes
firstly obeys the Stokes law $\nu \propto \zeta/\eta_0$ and secondly
is given by the Kramers' formula (\ref{eq3}) that is $k_K\propto \left (\zeta/\eta_0 \right )^{-1}$. Then the experimentally observable reaction rate constant is the average over the distribution
\begin{equation}
\label{eq12} k \left(\frac{\eta}{\eta_0 }\right)=\int\limits_{0}^{1}d q\ \rho (q)
k_K\left(\frac{\zeta}{\eta_0} \left(q, \frac{\eta}{\eta_0}\right)\right)
\end{equation}
The distribution $\rho (q)$ must be normalized
\begin{equation}
\label{eq13} \int\limits_{0}^{1}d q\ \rho (q) = 1
\end{equation}
As we want the reaction rate constant to be the experimentally observable (\ref{eq1}) and the local viscosity $\zeta$ to be (\ref{eq11})
then the relationship (\ref{eq12}) takes the form
\begin{equation}
\label{eq14} \frac{1}{\left (\eta/\eta_0 \right)^{\beta}}=\frac{1}{a}
\int\limits_{0}^{1}dq\ \frac{\rho(q)}
{(1-q)+q\left(\eta/\eta_0-1\right)}
\end{equation}
The latter is a Fredholm integral equation of the first kind for the unknown function of the distribution $\rho (q)$. It is analyzed in the next Sec.

\section{Analysis of the integral equation (\ref{eq14})}
We introduce the critical value of the weight $q$
\begin{equation}
\label{eq15} q_c=\frac{1}{2}
\end{equation}
Equation (\ref{eq14}) can be cast into the form solvable with the help of Fourier transform. Making use of N2.2.6.15. from \cite{Pru81} and the properties of the hypergeometric function
(see Appendix) one can easily verify by direct substitution into (\ref{eq14}) that in the range (\ref{eq5}) the normalized solution of (\ref{eq14}) is
\begin{equation}
\label{eq16} \rho (q)=\frac{\sin \left(\pi \beta \right)}{\pi q^{1-\beta}\left(q_c-q\right)^\beta}
\end{equation}
for $0 \leq q \leq q_c$ and
\begin{equation}
\label{eq17} \rho (q)=0
\end{equation}
for $q_c < q < 1$. At that the parameter $a$ and the the fractional power exponent $\beta$ must obey the relationship
\begin{equation}
\label{eq18} a=2^\beta
\end{equation}
From here we have
\begin{equation}
\label{eq19} \beta=\frac{ln\ a}{ln\ 2}
\end{equation}
In Fig. 1 the distribution $\rho(q)$ is depicted at several values of $\beta$.

\section{Relationship with experimental data}
The main experimental facts on the effect of solvent viscosity on the enzymatic reaction rate constant are: 1. the inverse proportionality to fractional power exponents of solvent viscosity given by
(\ref{eq1}) and 2. the existence of the limit value $\beta_{max}$ of the fractional power exponent with the decrease of cosolvent molecular weight given by (\ref{eq2}). The typical dependence of the fractional power exponent $\beta$ on the cosolvent molecular weight $M$ has the form \cite{Yed95}
\begin{equation}
\label{eq20} \beta \left( M\right)= \frac{b}{M^\delta}
\end{equation}
at $M_{min} \geq M \geq M_{max}$ so that $\beta_{max}=\beta \left( M_{min}\right)$.
The dependence (\ref{eq1}) is experimentally verified at values of solvent viscosity higher that some minimal value $\left (\eta/\eta_0\right)_{min}$.
For the oxygen escape from hemerythin the experimental values are: $\beta_{max}=0.79$,
$b=1.52$, $\delta=0.23$, $M_{min}=18$, $M_{max}=500000$ and $\left (\eta/\eta_0\right)_{min}\approx 1.6$ (see \cite{Yed95}).
Substitution of (\ref{eq20}) into (\ref{eq19}) yields the explicit dependence of the parameter $a$ on the cosolvent molecular weight
\begin{equation}
\label{eq21} a\left( M,b,\delta\right)=2^{b/M^\delta}
\end{equation}
In Fig.2 the dependence of the parameter $a\left( M,b,0.23\right)$ as the function of the logarithm of the cosolvent molecular weight $M$ is plotted for several values of the parameter $b$ at $\delta=0.23$. In Fig.3 the dependence of the parameter $a\left( M,1.52,\delta\right)$ as the function of the logarithm of the cosolvent molecular weight $M$ is plotted for several values of the parameter $\delta$ at $b=1.52$.
At $M \rightarrow M_{min}$ we have $a\left (M\right) \rightarrow a_{max}=a\left( M_{min},b,\delta\right)$.
The behavior of $a_{max}$ as the function of the parameters $b$ and $\delta$ is plotted in Fig.4.
In the limit $a \left( M,b,\delta\right)\rightarrow a_{max}$ the value of the fractional power exponent tends to the limit value $\beta_{max}$. From (\ref{eq19}) we have
\begin{equation}
\label{eq22} \beta_{max}=\frac{ln\ a_{max}}{ln\ 2}
\end{equation}
In Fig. 5 the fractional power exponent $\beta_{max}$ given by (\ref{eq22}) is plotted as the function of $a_{max}$. The experimental value $\beta_{max}=0.79$ for the oxygen escape from hemerythin from \cite{Yed95} is obtained at $a_{max}\approx 1.75$.

\section{Discussion}
A theoretical interpretation of the phenomenon of solvent viscosity dependence for the enzymatic reaction rate constant must account for two undisputable experimental facts: 1. the inverse proportionality to fractional power exponents of solvent viscosity given by (\ref{eq1}) and 2. the existence of the limit value $\beta_{max}$ of the fractional power exponent with the decrease of cosolvent molecular weight given by (\ref{eq2}). Up to now suggested models were mainly concerned with the explanation of the first fact while the second one was overlooked or ignored. On the other hand the model \cite{Sit08} suggests the explanation of the second fact but encounters difficulties at general description of the first one. The present approach is an alternative to all previous attempts that gives combined interpretation of both experimental facts in a self-consistent manner.

The basis of our approach is the relationship (\ref{eq8}). The latter is compatible with both nonlinear relationship of the viscosity for the movement of the system along the reaction coordinate with solvent viscosity (\ref{eq9}) and the linear one (\ref{eq11}). The choice among these options is by no means crucial for our approach. However we choose the linear relationship (\ref{eq11}) because in this case the appearance of the fractional dependence (\ref{eq1}) seems to be more nontrivial than in the case of nonlinear relationship. In the latter case we introduce in the model the fractional exponent $\kappa \not= 1$ from the very beginning by hands. This fact obscures the essence of our approach that in fact all additional fractional exponents and nonlinearities are superfluous.
In (\ref{eq8}) we factorize the viscosity for the movement of the system along the reaction coordinate $\zeta/\eta_0$ onto the contribution from protein matrix and solvent molecules and that from cosolvent molecules. The latter enters in $\zeta/\eta_0$ with the weight $q$. This weight may take different values from the range $0 < q < 1$ for the individual samples from the ensemble of enzymes. The ensemble is characterized by the distribution over the weight $\rho (q)$. We show that the very existence of this distribution is quite sufficient for the appearance of the required dependence of the reaction rate constant on solvent viscosity (\ref{eq1}) consistent with observations. Neither the modification of the Kramers' theory nor that of the Stokes law is in fact indispensable at such approach. In this regard our approach is conceptually much more simple than other interpretations of the phenomenon under consideration. The approach is motivated by vivid analogy of our problem with that of adsorption on heterogeneous surfaces and thus has apparent roots in the preceding experience of chemical physics. In this connection we note that we can cast (\ref{eq11}) into the form
\begin{equation}
\label{eq23}  \frac{\zeta}{\eta_0}=a\left [1+q\left(\frac{\eta}{\eta_0}-2\right)\right ]
\end{equation}
and call $q$ the "shielding parameter" that shows to what extent protein structure shields and screens the reaction coordinate from the solvent viscosity at $\eta/\eta_0 \geq 2$. This parameter is less than $1$ ($0 < q < 1$) because in experiment the solvent viscosity is varied by the cosolvent of the type of sugars, such as trehalose. It is known that "in trehalose solutions, there is generally a deficiency of trehalose and an excess of water in the vicinity of the protein" \cite{Bay03}. The deficiency of the cosolvent compared with the bulk in this case is by far takes place for protein interior because the larger the cosolvent molecular weight so much the worse it can be transmitted by protein structure to the reaction coordinate. Then the shielding parameter quantifies the measure of the deficiency of the cosolvent molecules in the vicinity of the reaction coordinate compared with the bulk.
We can represent the shielding parameter $q$ in the energetic form $\exp\left[-\epsilon /\left (k_BT\right )\right]$ where $\epsilon \geq 0$. In this case instead of the distribution $\rho (q)$ over the weight $q$ we obtain the one $\psi (\epsilon)$ over the energy $\epsilon $ that has the form identical to the well known Sips's distribution from the theory of adsorption on heterogeneous surfaces. However in our problem the physical meaning of the weight $q$ seems to be more lucid than that of the shielding parameter. That is why we prefer to deal with the former notion rather than with the latter one.

Our approach takes into account the experimental fact that the fractional power exponent $\beta$
tends to some limit value $\beta_{max}$ with the decrease of the cosolvent molecular weight $M$.
This fact is incorporated in the model as the inherent one.
The limit value $\beta_{max}$ is the function of the parameters $b$ and $\delta$. These parameters characterize the protein structure itself and its ability to transmit cosolvent molecules to the reaction coordinate. We conclude that the limit value of the fractional power exponent $\beta_{max}$ is a unique characteristic of the type of the enzyme (that determines the protein structure) rather than that of the solution of the enzymes. This conclusion is in agreement with experimental data (see Introduction).

The equation (\ref{eq17}) and \  Fig.1 show that the distribution $\rho (q)$ is zero above the critical value of the weight $q_c=0.5$ given by (\ref{eq15}). We conclude that the contribution from cosolvent molecules into the viscosity for the movement of the system along the reaction coordinate $\zeta/\eta_0$ can not enter with the weight greater that one half of that from protein matrix and solvent molecules to be compatible with observations expressed by (\ref{eq1}). For small values of the fractional power exponent $\beta$ the distribution is skewed to the upper bound $q=q_c$ of the range while for large values of $\beta$ it is skewed to the lower bound $q=0$.
In between $0 < q < q_c$ the distribution $\rho (q)$ tends to infinity at both $q \rightarrow 0+$ and $q \rightarrow q_c-$. Thus there are two main fractions of the samples in the ensemble of enzymes: those with pure contribution from protein matrix and solvent molecules (when $q \approx 0$) and those with the contribution from cosolvent molecules having the weight $q\approx q_c$ and the contribution from protein matrix and solvent molecules having the weight $\approx 1-q_c$. These fractions are the most representative in the ensemble of enzymes. However there are always fractions of the samples with intermediate values of the weight $q$ in between $0 < q < q_c$. The behavior of the distribution $\rho (q)$ in this region is quantified by the fractional power exponent $\beta$ (see Fig.1).
Thus the physical meaning of the fractional power exponent $\beta$ in the experimentally observable dependence of the enzymatic reaction rate constant on solvent viscosity (\ref{eq1}) is the characteristic of the behavior of the distribution $\rho(q)$ over the weight $q$ in the ensemble of enzymes. This distribution characterizes the solution of enzymes, i.e., is determined among with the type of the enzyme by such characteristics as the cosolvent molecular weight, pH, temperature, etc. The distribution  $\rho (q)$ over the weight $q$ acquires at our approach the status of the unique characteristic for the solution of enzymes.

We conclude that our approach yields conceptually simple interpretation and quantitative description of the main experimental data on solvent viscosity dependence for enzymatic reactions.

\section{Appendix}
The formula N2.2.6.15. from \cite{Pru81} is
\[
\int\limits_{0}^{a}d x\ x^{\alpha -1}(a-x)^{\beta-1}(x+z)^{-\rho}=
\]
\begin{equation}
\label{eq24}a^{\alpha+\beta-1}z^{-\rho}B(\alpha,\beta)
\ _2F_1\left(\alpha,\rho; \alpha+\beta; -a/z\right)
\end{equation}
where $\mid arg\ z \mid<\pi; a > 0, Re\ \alpha > 0, Re\ \beta > 0$.

The properties of the hypergeometric function used in the paper are:

see N6.7. in \cite{Pru03}

\begin{equation}
\label{eq25} _2F_1\left(a,\ b;\ c;\ z\right)=(1-z)^{-a}\ _2F_1\left(a,\ c-b;\ c;\ \frac{z}{z-1}\right)
\end{equation}

see N7.2.1.8. in \cite{Pru03}
\begin{equation}
\label{eq26}_2F_1\left(0,\ b;\ c;\ z\right)=\ _2F_1\left(a,\ 0;\ c;\ z\right)=1
\end{equation}

These properties are used at obtaining (\ref{eq16}).\\

Acknowledgements.  The author is grateful to Dr. Yu.F. Zuev for
helpful discussions. The work was supported by
the grant from RFBR and by the programme "Molecular and
Cellular Biology" of RAS.

\newpage

\begin{figure}
\begin{center}
\includegraphics* [width=\textwidth] {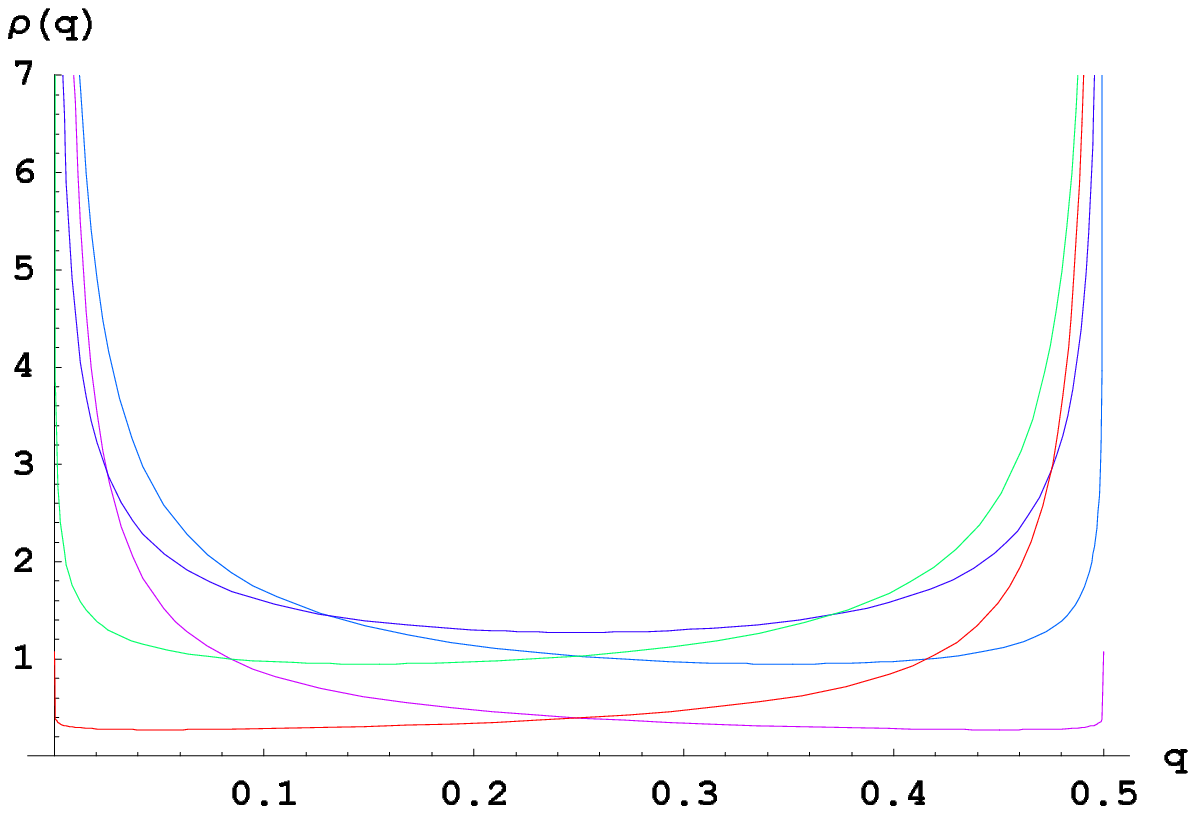}
\end{center}
\caption{Distribution $\rho (q)$ over the weight $q$ (eq. (\ref{eq16}) and eq. (\ref{eq17})) at increasing values of the fractional power exponent $\beta$: $\beta =0.1$ (purple color and the line is skewed to the upper bound $q=0.5$ of the range); $\beta =0.3$ (blue); $\beta =0.5$ (violet); $\beta =0.7$ (green); $\beta =0.8$ (red color and the line is skewed to the lower bound $q=0$ of the range).} \label{Fig.1}
\end{figure}

\clearpage
\begin{figure}
\begin{center}
\includegraphics* [width=\textwidth] {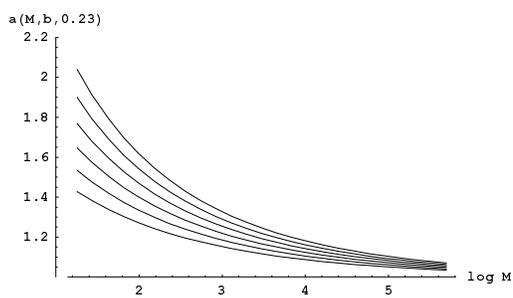}
\end{center}
\caption{The dependence of the parameter $a\left( M,b,\delta\right)$ (eq.(\ref{eq21})) as the function of the logarithm of the cosolvent molecular weight $M$ at $\delta=0.23$ for the values of the parameter $b$ from the down line to the upper one: $b=1$, $b=1.2$, $b=1.4$, $b=1.6$, $b=1.8$, $b=2$.}
\label{Fig.2}
\end{figure}

\clearpage
\begin{figure}
\begin{center}
\includegraphics* [width=\textwidth] {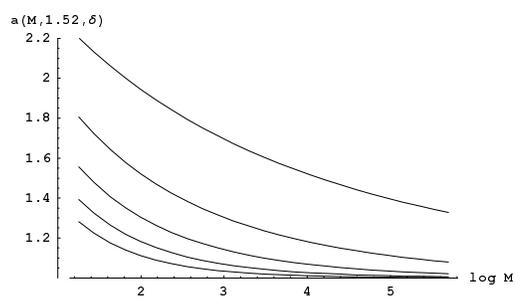}
\end{center}
\caption{The dependence of the parameter $a\left( M,b,\delta\right)$ (eq.(\ref{eq21})) as the function of the logarithm of the cosolvent molecular weight $M$ at $b=1.52$ for the values of the parameter $\delta$ from the down line to the upper one: $\delta=0.5$, $\delta=0.4$, $\delta=0.3$, $\delta=0.2$, $\delta=0.1$.}
\label{Fig.3}
\end{figure}

\clearpage
\begin{figure}
\begin{center}
\includegraphics* [width=\textwidth] {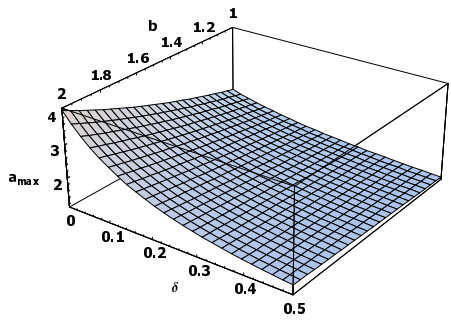}
\end{center}
\caption{The dependence of the parameter $a_{max}$ (eq. (\ref{eq21}) at $M_{min}=18$) as the function of the parameters $b$ and $\delta$. }
\label{Fig.4}
\end{figure}

\clearpage
\begin{figure}
\begin{center}
\includegraphics* [width=\textwidth] {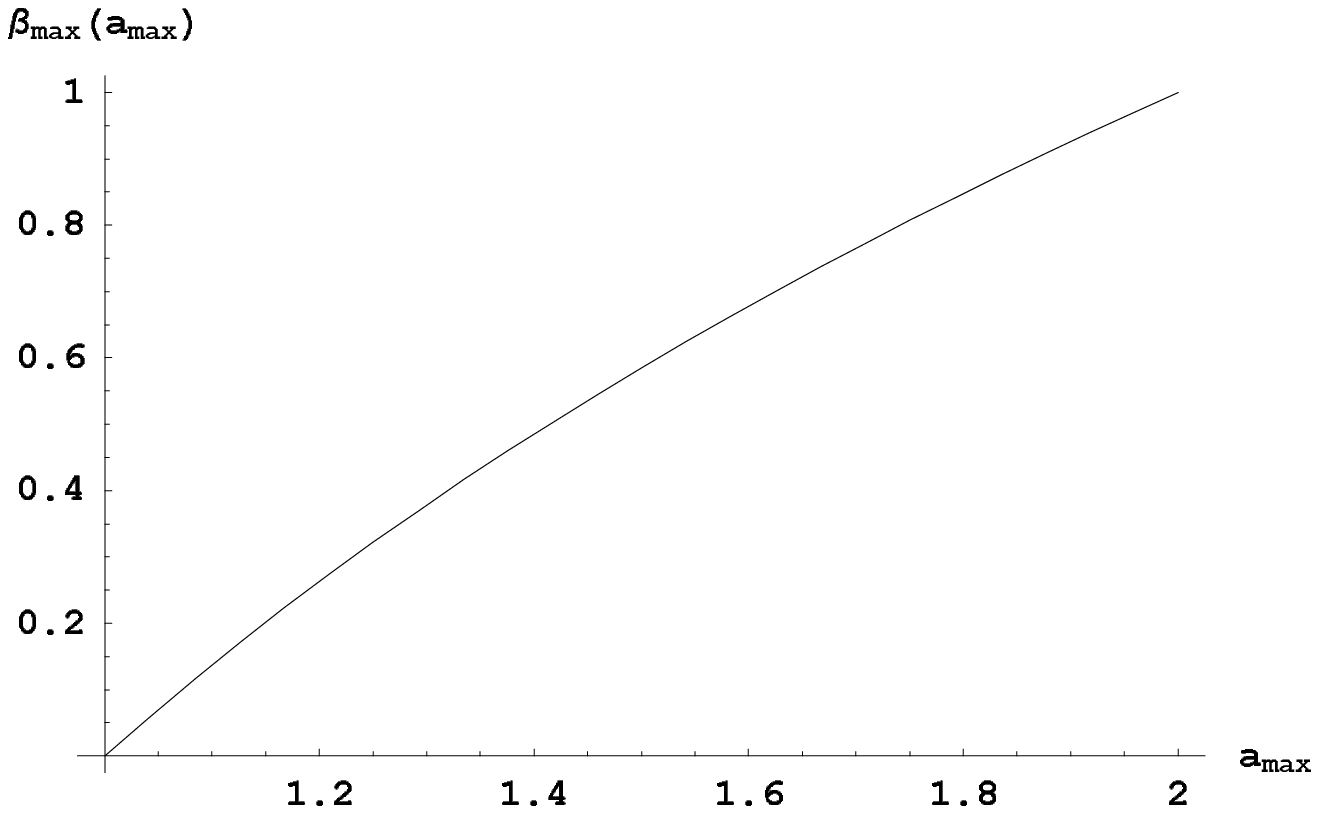}
\end{center}
\caption{The dependence of the limit value for the fractional power exponent $\beta_{max}$ (eq.(\ref{eq22})) as the function of the parameter $a_{max}$.}
\label{Fig.5}
\end{figure}

\newpage
\begin{thebibliography}{00}
\bibitem{Gav78}
B. Gavish, The role of geometry and elastic strains in
dynamic states of proteins, Biophys. Struc. Mech. 4 (1978) 37-52.
\bibitem{Gav79}
B. Gavish, M.M. Werber, Viscosity-Dependent Structural
Fluctuations in Enzyme Catalysis, Biochemistry 18 (1979) 1269-1275.
\bibitem{Bee80}
D. Beece, L. Eisenstein, H. Frauenfelder, D. Good,
 M. C. Marden, L. Reinisch, A. H. Reynolds, L. B. Sorensen
and K. T. Yue, Solvent Viscosity and Protein Dynamics, Biochemistry,
 19 (1980) 5147- 5157.
 \bibitem{Gav80}
B. Gavish, Position-dependent viscosity effects on rate coefficient,
Phys.Rev.Lett. 44 (1980) 1160-1163.
 \bibitem{Dos83}
W. Doster, Viscosity scaling and protein dynamics, Biophys. Chem. 17 (1983)
97-103.
 \bibitem{Gav86}
B. Gavish, Molecular dynamics and the transient strain model of enzyme
catalysis,
in:  C.R. Welsh, ed., The fluctuating enzyme, Wiley, N.Y., 1986.
\bibitem{Fra88}
H. Frauenfelder, F. Parak, R.D. Young, Conformational substates
in proteins, Ann.Rev.Biophys.Chem. 17 (1988) 451-479.
\bibitem{Dem89}
A.P. Demchenco, C.I. Rusyn, E.A. Saburova, Kinetics
of the lactate dehydrogenase reaction in high-viscosity media,
Biochem. et Biophys. Acta 998 (1989) 196-203.
\bibitem{Ng91}
K. Ng, A. Rosenberg,
Possible coupling of chemical to structural dynamics in subtilisin BPN'
catalyzed hydrolysis, Biophys Chem. 39 (1991) 57-68.
\bibitem{Ng911}
K. Ng, A. Rosenberg,
The coupling of catalytically relevant conformational fluctuations
 in subtilisin BPN' to solution viscosity revealed
by hydrogen isotope exchange and inhibitor binding,
Biophys Chem.  41 (1991) 289-99.
\bibitem{Gav94}
B. Gavish, S. Yedgar,
Solvent viscosity effects on protein dynamics: updating the concepts,
in: Protein-solvent interactions,
Ed. R. B. Gregory, Dekker, N. Y., 1994.
\bibitem{Dos94}
W. Doster, Th. Kleinert, F. Post, M. Settles,
Effect of solvent on protein internal dynamics,
in:  Protein-solvent interactions,
Ed. R. B. Gregory, Dekker, N. Y., 1994.
\bibitem{Yed95}
S. Yedgar, C. Tetreau, B. Gavish and D. Lavalette,
Viscosity dependence of $O_2$ escape from respiratory proteins as a function
of cosolvent molecular weight, Biophys. J. 68 (1995) 665-670.
\bibitem{Bar95}
G. Barshtein, A. Almagor, S. Yedgar, B. Gavish, Inhomogeneity of viscous
aqueous solutions, Phys. Rev. E52 (1995) 555-557.
\bibitem{Oh97}
H. Oh-oka, M. Iwaki, S. Itoh, Viscosity dependence of the electron transfer tate from bound cytochrome c to P840 in the photosynthetic reaction center of the green sulfur bacterium
Chlorobium tepidum, Biochemistry 36 (1997) 9267-9272.
\bibitem{Kle98}
Th. Kleinert, W. Doster, H. Leyser, W. Petry, V. Schwarz, M. Settles,
Solvent composition and viscosity effects on the kinetics of CO binding to horse
myoglobin, Biochemistry 37 (1998) 717-733.
\bibitem{Fra99}
H. Frauenfelder, P. G. Wolynes, R. H. Austin, Biological Physics,
Rev. Mod. Phys. 71 (1999) 419-430.
\bibitem{Uri03}
S. Uribe, J.G. Sampedro, Measuring solution viscosity and its effect on enzyme activity,
Biol. Proced. Online 5 (2003) 108-115.
\bibitem{Sit08}
A.E. Sitnitsky, Solvent viscosity dependence for enzymatic reactions,
Physica A 387 (2008) 5483-5497; arXiv:0804.2749v1[q-bio.BM].
\bibitem{Pab04}
S. A. Pabit, H. Roder, S. J. Hagen, Internal friction controls the speed
of protein folding from a
compact configuration, Biochemistry 43 (2004) 12532-12538.
\bibitem{Fra06}
H. Frauenfelder, P. W. Fenimore, G. Chen, B. H. McMahon,
Protein folding is slaved to solvent motions, Proc. Natl.
Acad. Sci. USA, 103 (2006) 15469-15472.
\bibitem{Kum08}
R. Kumar, A. K. Bhuyan, Viscosity scaling for the glassy phase of protein folding,
J. Phys. Chem. B  112 (2008) 12549–12554.
\bibitem{Jas01}
G. S. Jas, W. A. Eaton, J. Hofrichter, Effect of viscosity on the kinetics of $\alpha$-helix and $\beta$-hairpin formation, J. Phys. Chem. B  105 (2001) 261-272.
\bibitem{Han90}
P. H\"anggi, P. Talkner, M. Borkovec, Fifty years after Kramers'
equation: reaction rate theory, Rev.Mod.Phys. 62 (1990) 251-341.
\bibitem{Gro80}
R.F. Grote and J.T. Hynes, The stable states picture of chemical reactions.
II. Rate constants for condensed and gas phase reaction models, J.Chem.Phys.
73 (1980) 2715-2732.
\bibitem{Zwa92}
R. Zwanzig, Dynamical disorder: passage through a fluctuating bottleneck, J.
Chem. Phys. 97 (1992) 3587-3589.
\bibitem{Sit06}
A.E. Sitnitsky, Dynamical contribution into enzyme catalytic
efficiency, Physica A371 (2006) 481-491; arXiv:cond-mat/0601165.
\bibitem{Sch88}
J. Schlitter, Viscosity dependence of intramolecular activated processes,
Chemical Physics 120 (1988) 187-197.
\bibitem{Sum91}
H. Sumi, Theory on reaction rates in nonthermallzed steady states during conformational
fluctuations in viscous solvents, J. Phys. Chem. 95 (1991) 3334-3350.
\bibitem{Rud92}
W. Rudzihski, D.H. Everett, Adsorption of Gases on Heterogeneous
Surfaces, Academic, London, 1992.
\bibitem{Cer93}
G.F. Cerofolini, N. Re, The mathematical theory of adsorption
on non-ideal surfaces, Rivista Del Nuova Cimento 16 (1993) 1-63.
\bibitem{Zel34}
Ya.B. Zel'dovich, To the theory of Freundlich adsorption isotherm,
Acta physicochemica URSS 1 (1934) 961-974.
\bibitem{Tho67}
J.M. Thomas, W.J. Thomas, Introduction to the principles of heterogeneous
 catalysis, N.Y. Acad.Press,  1967, pp. 43-45.
\bibitem{Sip48}
R. Sips, On the structure of a catalyst surface, J. Chem. Phys. 16 (1948)
490-495.
\bibitem{Sip50}
R. Sips, On the structure of a catalyst surface II, J. Chem. Phys. 18 (1950)
1024-1026.
\bibitem{Mis70}
D.N. Misra, New adsorption isotherm for heterogeneous surfaces,
J. Chem. Phys. 52 (1970) 5499-5501.
\bibitem{Cer72}
G.F. Cerofolini, Multilayer adsorption on heterogeneous surfaces,
Journal of Low Temperature Physics, 6 (1972) 473-486.
\bibitem{Rud75}
W. Rudzihski, M. Jaroniec, S. Sokotowski, G.F. Cerofolini,
Gas adsorption on heterogeneous surfaces: a detailed computation of
adsorption energy distribution, Czech. J. Phys. B 25 (1975) 891-901.
\bibitem{Jar75}
M. Jaroniec, W. Rudzihski, S. Sokotowski, R. Smarzewski, Determination
of energy distribution function from observed adsorption isotherms,
Colloid and Polymer Sci. 253 (1975) 164-166.
\bibitem{Lan76}
U. Landman, E.W. Montroll, Adsorption on heterogeneous surfaces. I.
Evaluation of the energy distribution function via the Wiener and Hopf
method, J. Chem. Phys. 64 (1976) 1762-1767.
\bibitem{Mag04}
S. Magazu, G. Maisano, F. Migliardo, C. Mondelli, Mean-square displacement relationship in bioprotectant systems by elastic neutron scattering, Biophysical Journal 86 (2004) 3241–3249.
\bibitem{Cor05}
E. Cornicchi, G. Onori, A. Paciaroni, Picosecond-time-scale fluctuations of proteins in glassy matrices: the role of viscosity, Phys.Rev.Lett. 95 (2005) 158104.
\bibitem{McC77}
Y.A. McCammon, P. Wolynes, Nonsteady hydrodynamics of biopolymer motions,
J. Chem. Phys. 66 (1977) 1452-1456.
\bibitem{Kna82}
E.W. Knapp, S.F. Fisher, F. Parak, Protein dynamics from Moessbauer spectra. The temperature dependence, J. Phys. Chem. 86 (1982) 5042-5047.
\bibitem{Pru81}
A.P. Prudnikov, Yu.A. Brychkov, O.I. Marichev, Integrals and serieses.
Elementary functions. V.1, Nauka, Moscow, 1981.
\bibitem{Bay03}
B.M. Baynes,  B.L. Trout, Proteins in Mixed Solvents: A Molecular-Level Perspective,
J. Phys. Chem. B 107 (2003) 14058-14067.
\bibitem{Pru03}
A.P. Prudnikov, Yu.A. Brychkov, O.I. Marichev, Integrals and serieses.
Special functions. Additional chapters. V.3, 2-nd ed., Fizmatlit, Moscow, 2003.
\end{thebibliography}
\end{document}